\documentstyle[12pt]{article}
\catcode`\@=11

\@addtoreset{equation}{section}
\def\theequation{\arabic{section}.\arabic{equation}}

\def\@normalsize{\@setsize\normalsize{15pt}\xiipt\@xiipt
\abovedisplayskip 14pt plus3pt minus3pt%
\belowdisplayskip \abovedisplayskip
\abovedisplayshortskip  \z@ plus3pt%
\belowdisplayshortskip  7pt plus3.5pt minus0pt}

\def\small{\@setsize\small{13.6pt}\xipt\@xipt
\abovedisplayskip 13pt plus3pt minus3pt%
\belowdisplayskip \abovedisplayskip
\abovedisplayshortskip  \z@ plus3pt%
\belowdisplayshortskip  7pt plus3.5pt minus0pt
\def\@listi{\parsep 4.5pt plus 2pt minus 1pt
            \itemsep \parsep
            \topsep 9pt plus 3pt minus 3pt}}

\def\underline#1{\relax\ifmmode\@@underline#1\else
        $\@@underline{\hbox{#1}}$\relax\fi}
\@twosidetrue

\relax

\catcode`@=12

\catcode`\@=11

\def\section{\@startsection{section}{1}{\z@}{3.5ex plus 1ex minus
   .2ex}{2.3ex plus .2ex}{\large\bf}}

\def\thesection{\Roman{section}.}

\def\appendix{\setcounter{section}{0}
        \def\thesection{APPENDIX }
        \def\theequation{\Alph{section}.\arabic{equation}}}

\def\FERMIPUB{}

\def\ps@headings{\def\@oddfoot{}\def\@evenfoot{}
\def\@oddhead{\hbox{}\hfill
        \makebox[.5\textwidth]{\raggedright\ignorespaces --\thepage{}--
        \hfill {\rm FERMILAB--Pub--\FERMIPUB}}}
\def\@evenhead{\@oddhead}
\def\subsectionmark##1{\markboth{##1}{}}
}

\catcode`\@=12

\relax

\def\figcap{\section*{Figure Captions\markboth
        {FIGURECAPTIONS}{FIGURECAPTIONS}}\list
        {Fig. \arabic{enumi}:\hfill}{\settowidth\labelwidth{Fig. 999:}
        \leftmargin\labelwidth
        \advance\leftmargin\labelsep\usecounter{enumi}}}
 \relax
\def\tablecap{\section*{Table Captions\markboth
        {TABLECAPTIONS}{TABLECAPTIONS}}\list
        {Table \arabic{enumi}:\hfill}{\settowidth\labelwidth{Table 999:}
        \leftmargin\labelwidth
        \advance\leftmargin\labelsep\usecounter{enumi}}}
 \relax
\def\reflist{\section*{References\markboth
        {REFLIST}{REFLIST}}\list
        {[\arabic{enumi}]\hfill}{\settowidth\labelwidth{[999]}
        \leftmargin\labelwidth
        \advance\leftmargin\labelsep\usecounter{enumi}}}
 \relax

\catcode`\@=11

\def\FERMIPUB{}

\def\ps@headings{\def\@oddfoot{}\def\@evenfoot{}
\def\@oddhead{\hbox{}\hfill
        \makebox[.5\textwidth]{\raggedright\ignorespaces --\thepage{}--
        \hfill {\rm FERMILAB--Pub--\FERMIPUB}}}
\def\@evenhead{\@oddhead}
\def\subsectionmark##1{\markboth{##1}{}}
}

\ps@headings

\relax
\newskip\humongous \humongous=0pt plus 1000pt minus 1000pt
\def\caja{\mathsurround=0pt}
\def\eqalign#1{\,\vcenter{\openup1\jot \caja
        \ialign{\strut \hfil$\displaystyle{##}$&$
        \displaystyle{{}##}$\hfil\crcr#1\crcr}}\,}
\newif\ifdtup

\def\beq{\begin{equation}}
\def\eeq{\end{equation}}

\def\beqn{\begin{eqnarray}}
\def\eeqn{\end{eqnarray}}
\relax

\def\G2{{\; \rm GeV/}c^2}
\def\G{\; \rm GeV}

\def\dotx{\dotx{\dot\overline{x}}}

\relax

\def\conj{\bigcirc \hskip -.24cm *}
\begin{document}
\hbadness=10000
\begin{titlepage}
\nopagebreak
\begin{flushright}

        {\normalsize KUCP-52\\

        October,~1992}\\
\end{flushright}
\vfill
\begin{center}
{\large \bf q-Deformed Conformal and Poincar{\'e} \\ Algebras
on Quantum 4-spinors}
\vfill
\renewcommand{\thefootnote}{\fnsymbol{footnote}}
{\bf Tatsuo Kobayashi\footnote{
Fellow of the Japan Society for the Promotion of Science. Work
partially supported by the Grant-in-Aid for Scientific Research from the
Ministry of Education, Science and Culture (\# 030083)} and Tsuneo Uematsu
\footnote{Work partially supported by the Grant-in-Aid for Scientific Research
from the Ministry of Education, Science and Culture (\# 04245221)}}

       Department of Fundamental Sciences, FIHS, \\
       Kyoto University,~Kyoto 606,~Japan \\
\vfill

\end{center}

\vfill
\nopagebreak
\begin{abstract}
We investigate quantum deformation of conformal algebras by constructing
the quantum space for $sl_q(4,C)$. The differential calculus
on the quantum space and the action of the quantum generators are studied.
We derive deformed $su(4)$ and $su(2,2)$ algebras from the deformed $sl(4)$
algebra using the quantum 4-spinor and its conjugate spinor.
The 6-vector in $so_q(4,2)$ is constructed as a tensor product of two sets
of 4-spinors.  The reality condition for the 6-vector and that for the
generators are found.  The q-deformed Poincar{\'e} algebra is extracted
as a closed subalgebra.

\end{abstract}

\vfill
\end{titlepage}
\pagestyle{plain}
\newpage
\voffset = -2.5 cm
\leftline{\large \bf 1. Introduction}
\vspace{0.8 cm}

Recently much attention has been paid on quantum groups and quantum
algebras \cite{Drinfeld,Jimbo,Faddeev-Reshetikhin-Takhtajan,KR} in both
theoretical physics and mathematics \cite{Doebner-Hennig,CFZ,Takhtajan,K}.
So far most of the applications are restricted to the Lie groups or algebras
corresponding to the internal symmetries.  Now it would be intriguing to
investigate possible quantum deformations of fundamental space-time
symmetries, which might be relevant at very short distance
e.g. at Planck length implied by the unification of basic interactions
including gravity.
In the last few years, there have been pioneering works in extending
quantum-group ideas to space-time symmetries including Lorentz, Poincar{\'e}
and
conformal algebras
\cite{PW,CSS-Watamura,SWZ,OSWZ2,OSWZ,Lukierski-Ruegg-Nowicki-Tolstoy,LN,LNR}.

The quantum group deformation can be realized on the quantum space or quantum
(hyper-)plane in which the coordinates are non-commutative
\cite{Manin,Takhtajan}.
The differential calculus on the non-commutative space of quantum groups
has been explored by Woronowicz \cite{Woronowicz}, which provides us with an
example for noncommutative differential geometry \cite{Connes}.
Wess and Zumino and other people developed the differential calculus on the
quantum (hyper-)plane covariant with respect to quantum groups
\cite{Wess-Zumino} and multiparameter deformation of the quantum groups,
especially for $GL_q(n)$
\cite{Schirrmacher-Wess-Zumino,Schirrmacher1,Schirrmacher2},
as well as for $SO_q(n)$ \cite{Carow-Watamura-Schlieker-Watamura}.
Based on this framework quantum deformations of Lorentz group and algebra
\cite{PW,CSS-Watamura,SWZ,OSWZ2}
have been studied by deforming $SL(2,C)$.  The q-deformed
Poincar{\'e} algebra has also been studied along this line \cite{OSWZ}.
In Ref. \cite{LN,LNR}, deformation of the conformal
algebra has been discussed in Drinfeld-Jimbo procedure.
In additon to the algebra, a space representing the algebra is interesting.
Therefore, we here study the deformed space-time reprsenting the deformed
conformal algebra, as well as the deformed conformal algebra.

In this paper we investigate q-deformation of $D=4$ conformal algebra based
on the quantum space which realizes $SL_q(4,C)$ where the quantum 4-spinors
appears as basic ingredients for which we set up commutation relations and
study the differential calculus.  We then investigate the action of quantum
generators of $sl_q(4,C)$ on these quantum 4-spinors.
We next introduce quantum conjugate 4-spinors, in order to derive a charge
conjugation of the generators and to obtain deformed $su(4)$ and $su(2,2)$
algebras.
The quantum 6-vector in $so_q(4,2)$ is constructed as a bi-spinor of two
sets of 4-spinors, and the conjugation of the 6-vector which guarantees the
reality of the 6-vector is presented.  For a suitable conjugation one can
obtain the q-deformed Poincar{\'e} algebra as a closed subalgebra of
q-deformed conformal algebra.

In the next section we introduce the quantum space realizing $SL_q(4,C)$
and study the differential calculus on the quantum 4-spinor space.  From the
consistency conditions we derive the action of the generators on quantum
4-spinors.  In section 3 we study conjugate spinors and q-deformed $su(4)$
algebra.
We next examine the quantum 6-vector constucted as the bilinear combination
of two 4-spinors in section 4.  In section 5 we study the reality conditions
for the 6-vector as well as for the generators and present q-deformed
conformal
algebra. In section 6 we extract a q-deformation
of Poincar{\'e} as a subalgebra of q-deformed conformal algebra
and construct a quantum Casimir invariant.  The final section
will be devoted to some concluding remarks.

\vspace{0.8 cm}
\leftline{\large \bf 2. Deformed $sl(4)$ algebra on quantum space}
\vspace{0.8 cm}

We set up the commutation relations for the
coordinates $x^i$ and derivatives $\partial_i$ on the quantum space
\cite{Wess-Zumino} as follows,
$$ x^i x^j =q x^j x^i, \qquad \partial_i \partial_j ={1 \over q} \partial_j
\partial_i,$$
$$ \partial_i x^k=q x^k \partial_i,      \eqno (2.1)$$
$$ \partial_i x^i=1+q^2x^i \partial_i +(q^2-1)\sum_{j>i}x^j \partial_j,$$
where $i<j$.
These relations are governed by a $\hat R$-matrix of $SL_q(n)$ quantum group,
which is explicitly given as
$$ \hat R^{ij}_{\ \ k\ell}= \delta^i_{\ \ell}\delta^j_{\ k}((1-q^{-1})
\delta^{ij}+q^{-1})+(1-q^{-2})\delta^i_k \delta^j_\ell \Theta^{ji},
\eqno(2.2)$$
where $\Theta^{ij}$ is equal to 1 for $i>j$, otherwise vanishes.
For example, the commutation relation of coordinates is written in terms of
the $\hat R$-matrix as $x^ix^j=\hat R^{ij}_{\ \ k \ell}x^k x^\ell$.
The $\hat R$-matrix is decomposed into two projection operators, i.e., a
symmetric one $\cal S$ and an antisymmetric one $\cal A$ as
$$ {\cal S}={1 \over 1+q^{-2}}(\hat R +q^{-2}{\bf 1}), \qquad {\cal A}=
{-1 \over 1+q^{-2}}(\hat R -{\bf 1}). \eqno(2.3)$$

Now, let us study deformed $sl(4)$ algebra on the 4-dim quantum space, which
is also called a \lq\lq quantum 4-spinor " hereafter.
First of all, we consider actions of generators $T^\ell_{\ \ell+1}$
($\ell =1 \sim 3$) which correspond to $x^\ell \partial_{\ell+1}$ and
associate with simple roots in the classical limit ($q \rightarrow 1$).
Following Ref.\cite{SWZ}, we assume the actions of
$T^\ell_{\ \ell+1}$ on $x^i$ and $\partial_i$ as
$$ T^\ell_{\ \ell+1} x^i=a^{(\ell)i}x^i T^\ell_{\ \ell +1}+
\delta^i_{\ell +1}x^\ell, \eqno(2.4.a)$$
$$ T^\ell_{\ \ell+1} \partial_i=b^{(\ell)}_i \partial_i T^\ell_{\ \ell +1}+
\beta^{(\ell)}\delta^{\ell}_i\partial_{\ell+1}. \eqno(2.4.b)$$
Eqs. (2.4) should be consistent with the commutation relations (2.1).
We calculate $T^\ell_{\ \ell+1}(x^ix^j-qx^jx^i)$ ($i<j$) so as to derive
consistency conditions as follows,
$$ \delta^i_{\ell+1}(x^\ell x^j-qa^{(\ell)j}x^j x^\ell)+\delta^j_{\ell+1}
(a^{(\ell)i}x^i x^\ell-qx^\ell x^i)=0. \eqno (2.5)$$
Namely, we obtain $a^{(\ell)\ell}=q$ and $a^{(\ell)j}=1$ for $j \neq \ell,
\ell+1$.
It is remarkable that if we only consider the action of the generators on
$x^i$, we are not able to determine coefficients of (2.4.a) completely, i.e.,
$a^{(\ell)\ell+1}$ is undetermined.
Consistency between eq. (2.4.b) and the commutation relation $\partial_i
\partial_j-q\partial_j \partial_i=0$ provides $b^{(\ell)}_{\ell+1}=q$ and
$b^{(\ell)}_\ell=1$ for $i \neq \ell, \ell+1$.
Further, consistency between eqs. (2.4) and the commutaion relations of $x$
and $\partial$
leads to $a^{(\ell)\ell+1}=1/q$, $b^{(\ell)}_\ell=1/q$ and
$\beta^{(\ell)}=-1/q$.
Thus we obtain the action of $T^\ell_{\ \ell+1}$ on $x$ and $\partial$ as
follows,
$$ [T^\ell_{\ \ell+1},x^i]=[T^\ell_{\ \ell+1},\partial_i]=0,$$
$$T^\ell_{\ \ell+1} x^\ell=qx^\ell T^\ell_{\ \ell+1},$$
$$ T^\ell_{\ \ell+1} x^{\ell+1}=q^{-1}x^{\ell+1}T^\ell_{\ \ell+1}+ x^\ell,
\eqno (2.6)$$
$$ T^\ell_{\ \ell+1} \partial_\ell =q^{-1} \partial_\ell T^\ell_{\ \ell+1}
-q^{-1} \partial_{\ell+1},$$
$$ T^\ell_{\ \ell+1}\partial_{\ell+1} = q\partial_{\ell+1}T^\ell_{\ \ell+1},$$
where $i \neq \ell, \ell+1$.
These relations show that $T^1_{\ 2}$ and $T^3_{\ 4}$ commute with each other.

Next, we define $T^\ell_{\ \ell+2}$ ($\ell=1,2$) as
$$ T^\ell_{\ \ell+2} \equiv [T^\ell_{\ \ell+1},T^{\ell+1}_{\ \ell+2}]_{f^\ell}
, \eqno(2.7)$$
where $[A,B]_p \equiv AB-pBA$.
Using (2.6), we can easily obtain actions of $T^\ell_{\ \ell+2}$ on $x$ and
$\partial$.
In terms of $T^\ell_{\ \ell+2}$, we can define two generators corresponding
to $x^1 \partial_4$ in the classical limit as follows,
$$ T^1_{\ 4} \equiv [T^1_{\ 3},T^3_{\ 4}]_g, \qquad T'^1_4 \equiv [T^1_{\ 2},
T^2_{\ 4}]_{g'}. \eqno(2.8)$$
They act on $x^i$ as follows,
$$ T^1_{\ 4} x^1 =q x^1 T^1_{\ 4} ,$$
$$ T^1_{\ 4} x^2 =x^2 T^1_{\ 4} +(q-f^1)x^1 [T^2_{\ 3},T^3_{\ 4}]_g, $$
$$ T^1_{\ 4} x^3=x^3T^1_{\ 4} + (1-qf^1)(1-{g \over q})x^2 T^1_{\ 2}
T^3_{\ 4}+(q-g)x^1 T^3_{\ 4},$$
$$ T^1_{\ 4} x^4=q^{-1}x^4 T^1_{\ 4} +(q^{-1}-g)x^3 T^1_{\ 3}+
(q^{-1} -f^1)x^2 T^1_{\ 2}+x^1,$$
$$ T'^1_{\ 4} x^1 =q x^1 T'^1_{\ 4} , \eqno(2.9)$$
$$ T'^1_{\ 4} x^2 =x^2 T'^1_{\ 4} +(q-g')x^1 T^2_{\ 4}, $$
$$ T'^1_{\ 4} x^3=x^3T'^1_{\ 4} + (1-qg')(1-{f^2 \over q})x^2 T^1_{\ 2}
T^3_{\ 4}+(q-f^2)x^1 T^3_{\ 4},$$
$$ T'^1_{\ 4} x^4=q^{-1}x^4 T'^1_{\ 4} +(q^{-1}-f^2)x^3
[T^1_{\ 2},T^2_{\ 3}]_{g'}+(q^{-1} -g')x^2 T^1_{\ 2}+x^1.$$
The coefficients should satisfy conditions that $g=f^2$ and $g'=f^1$ so that
$T^1_{\ 4}$ and $T'^1_{\ 4}$ are identified.
Further, the coefficients should obey a condition that $f^1=1/q$ or $f^2=q$
so that the bilinear term $T^1_{\ 2}T^3_{\ 4}$ vanishes on the right hand
side of eq. (2.9).
Under the above conditions, actions of the two generators $T^1_{\ 4}$ and
$T'^1_{\ 4}$ on $\partial_i$ are also identified and do not have bilinear
terms of $T$.

Next, we have to investigate closure of the algebra of the generators
$T^i_{\ j}$ ($i<j$).
For a concrete example, let us consider a commutation relation between
$T^\ell_{\ \ell+1}$ and $T^{\ell}_{ \ \ell+2}$.
Actions of $T^\ell_{\ \ell+1}T^{\ell}_{ \ \ell+2}$ and $T^{\ell}_{\ \ell+2}
T^\ell_{\ \ell+1}$ on $x$ are obtained as
$$ [T^\ell_{\ \ell+1} T^\ell_{\ \ell+2},x^i]=[T^\ell_{\ \ell+2}
T^\ell_{\ \ell+1},x^i]=0,\qquad (i=\ell-1 {\rm \ or \ } \ell+3),$$
$$ T^\ell_{\ \ell+1} T^\ell_{\ \ell+2} x^\ell=q^2 x^\ell T^\ell_{\ \ell+1}
T^\ell_{\ \ell+2},$$
$$ T^\ell_{\ \ell+1} T^\ell_{\ \ell+2} x^{\ell+1}=q^{-1}x^{\ell +1}
T^\ell_{\ \ell +1}T^\ell_{\ \ell+2}+x^\ell T^\ell _{\ \ell+2}+
q(q-f^\ell)x^\ell T^\ell_{\ \ell +1} T^{\ell +1}_{\ \ell+2},$$
$$\eqalign{ T^\ell_{\ \ell+1} T^\ell_{\ \ell+2} x^{\ell+2}= &
q^{-1}x^{\ell+2} T^\ell_{\ \ell+1} T^\ell_{\ \ell+2} +q^{- 1}
(q^{-1}-f^\ell)x^{\ell +1}T^\ell_{\ \ell+1}T^\ell_{\ \ell +1} \cr
& +(q+q^{-1}-f^\ell)x^\ell T^\ell_{\ \ell+1},}$$
$$ T^\ell_{\ \ell+2} T^\ell_{\ \ell+1} x^\ell=q^2 x^\ell T^\ell_{\ \ell+2}
T^\ell_{\ \ell+1}, \eqno (2.10)$$
$$ T^\ell_{\ \ell+2} T^\ell_{\ \ell+1} x^{\ell+1}=q^{-1}x^{\ell +1}
T^\ell_{\ \ell +2}T^\ell_{\ \ell+1}+qx^\ell T^\ell _{\ \ell+2}+
q^{-1}(q-f^\ell)x^\ell T^{\ell +1}_{\ \ell +2} T^\ell_{\ \ell+1},$$
$$ T^\ell_{\ \ell+2} T^\ell_{\ \ell+1} x^{\ell+2}=q^{-1}x^{\ell+2}
T^\ell_{\ \ell+2} T^\ell_{\ \ell+1} +(q^{-1}-f^\ell)x^{\ell +1}
T^\ell_{\ \ell+1}T^\ell_{\ \ell +1}+x^\ell T^\ell_{\ \ell+1}.$$
Eq. (2.10) shows that if $f^\ell$ is equal to $q$ or $1/q$, $f^\ell
T^\ell_{\ \ell +1} T^\ell_{\ \ell+2}$ is identified with
$T^\ell_{\ \ell +2}T^\ell_{\ \ell +1}$.
Similarly the algebra of the other generators close only if
$f^\ell=q$ or $1/q$.
Here, we choose a basis of the algebra where $f^1=f^2=q$.
Under the basis, we can summarize the actions of $T^\ell_{\ k}$ ($\ell<k$) as
follows,
$$ T^\ell_{\ k} x^\ell= q x^\ell T^\ell_{\ k},$$
$$ T^\ell_{\ k} x^k = q^{-1}x^kT^\ell_{\ k}+x^\ell+(q^{-1}-q)
\sum^{k-1}_{j=\ell+1}x^j T^\ell_{\ j},$$
$$ T^\ell_{\ k} x^i=x^iT^\ell_{\ k},  \qquad (i\neq \ell, k), \eqno (2.11)$$
$$ T^\ell_{\ k} \partial_i= \partial_i T^\ell_{\ k}, \qquad
(i<\ell \ {\rm or} \ k<i),$$
$$ T^\ell_{\ k} \partial_\ell = q^{-1} \partial_\ell T^\ell_{\ k} - q^{-1}
\partial_k,$$
$$T^\ell_{\ k} \partial _k= q \partial_k T^\ell_{\ k},$$
$$T^\ell_{\ k} \partial _j = \partial _jT^\ell_{\ k} +(q-q^{-1})\partial_k
T^\ell_{\ j},\qquad (\ell<j<k).$$
Futher, the algebra of these generators is obtained as follows,
$$[T^\ell_{\ j},T^\ell_{\ i}]_q=[T^j_{\ k},T^i_{\ k}]_q=0, \qquad (i<j), $$
$$[T^1_{\ 2},T^3_{\ 4}]=[T^1_{\ 4},T^2_{\ 3}]=0, $$
$$ [T^\ell_{\ j},T^j_{\ k}]_q=T^\ell_{\ k}, \eqno (2.12)$$
$$ [T^2_{\ 4},T^1_{\ 3}]=(q-q^{-1})T^1_{\ 4}T^2_{\ 3}.$$
Similarly we can obtain actions of $T^k_{\ \ell}$ ($\ell < k$), and algebra
of them.
The results are summarized in Appendix.

Now let us define three Cartan elements $H_\ell$ ($\ell=1 \sim 3$) in terms
of commutation relations between $T^{\ell +1}_{\ \ell}$ and
$T^\ell_{\ \ell+1}$.
Making use of (2.11) and (A.1), we can easily find actions of
$T^{\ell +1}_{\ \ell} T^\ell_{\ \ell +1}$ and $T^{\ell}_{\ \ell+1}
T^{\ell+1}_{\ \ell}$ on $x^i$ as follows,
$$T^{\ell}_{\ \ell+1} T^{\ell +1}_{\ \ell} x^\ell= q^2 x^\ell
T^{\ell}_{\ \ell +1} T^{\ell +1}_{\ \ell}+q^{-1} x^{\ell +1}
T^\ell_{\ \ell+1}+x^\ell,$$
$$T^{\ell}_{\ \ell+1} T^{\ell +1}_{\ \ell} x^{\ell +1}= q^{-2}
x^{\ell +1}T^{\ell}_{\ \ell+1} T^{\ell +1}_{\ \ell}+q^{-1} x^{\ell }
T^{\ell +1}_{\ \ell}, \eqno (2.13)$$
$$T^{\ell +1}_{\ \ell} T^\ell_{\ \ell +1} x^\ell= q^2 x^\ell
T^{\ell +1}_{\ \ell} T^\ell_{\ \ell +1}+q x^{\ell +1}T^\ell_{\ \ell+1},$$
$$T^{\ell +1}_{\ \ell} T^\ell_{\ \ell +1} x^{\ell +1}= q^{-2} x^{\ell +1}
T^{\ell +1}_{\ \ell} T^\ell_{\ \ell +1}+q x^{\ell }T^{\ell +1}_{\ \ell} +
x^{\ell +1}.$$
Following Ref.\cite{SWZ}, we define $H_\ell$ by linear
combination of $T^\ell_{\ \ell+1} T^{\ell +1}_{\ \ell}$ and
$T^{\ell +1}_{\ \ell}T^{\ell}_{\ \ell +1}$ in order to eliminate the linear
term of $T$ on the right hand side of (2.13
$$ H_\ell \equiv q^{-1} T^{\ell +1}_{\ \ell} T^\ell_{\ \ell +1}-q
T^\ell_{\ \ell +1} T^{\ell+1}_{\ \ell}. \eqno (2.14)$$
They act on $x$ and $\partial$ as
$$ H_\ell x^\ell = q^2 x^\ell H_\ell - q x^\ell,$$
$$ H_\ell x^{\ell +1} = q^{-2} x^{\ell +1} H_\ell + q^{-1} x^{\ell +1},$$
$$ H_\ell \partial_\ell = q^{-2} \partial_\ell H_\ell +q^{-1}\partial_\ell,
\eqno (2.15)$$
$$ H_\ell \partial_{\ell +1} = q^2 \partial_{\ell +1} H_\ell -q
\partial _{\ell +1},$$
$$ [H_\ell,x^i]=[H_\ell,\partial_i]=0, \qquad (i\neq \ell, \ell+1).$$
Using (2.15), we can easily find that the Cartan elements $H_\ell$ commute
with each other, i.e.,
$$ [ H_\ell,H_k]=0. \eqno (2.16)$$
Making use of (2.11) and (A.1), we can calculate commutation relations
between $T^\ell_{\ \ell+1}$ and $T^{k+1}_{\ k}$ ($\ell \neq k$) as follows,
$$ [T^\ell_{\ \ell+1},T^{\ell +2}_{\ \ell+1}]_q=[T^{\ell +1}_{\ \ell +2},
T^{\ell+1}_{\ \ell}]_q=0,$$
$$ [T^1_{\ 2},T^4_{\ 3}]=[T^3_{\ 4},T^2_{\ 1}]=0. \eqno (2.17)$$
Moreover, we can obtain commutation relations between $H_\ell$ and
$T^j_{\ k}$ for $k= j +1$ or $j -1$ as follows,
$$ q^{-2} H^\ell T^\ell_{\ \ell+1}-q^2T^\ell_{\ \ell +1}H^\ell =-(q+q^{-1})
T^\ell_{\ \ell +1},$$
$$ q H^\ell T^{\ell +1}_{\ \ell +2}-q^{-1}T^{\ell +1}_{\ \ell +2}
H^\ell=T^{\ell +1}_{ \ \ell +2},$$
$$ qH^{\ell +1}T^\ell_{\ \ell +1}-q^{-1}T^\ell_{\ \ell+1}H^{\ell +1}=
T^\ell_{\ \ell +1},$$
$$ q^2H^\ell T^{\ell +1}_{\ \ell}-q^{-2}T^{\ell +1}_{\ \ell}
H^\ell=(q+q^{-1})T^{\ell +1}_{\ \ell}, \eqno (2.18)$$
$$ q^{-1}H^\ell T^{\ell +2}_{\ \ell +1}-qT^{\ell +2}_{\ \ell+ 1}
H^\ell=-T^{\ell+2}_{\ \ell +1},$$
$$ q^{-1}H^{\ell +1} T^{\ell +1}_{\ \ell}-qT^{\ell +1}_{\ \ell}H^{\ell +1}
=-T^{\ell+1}_{\ \ell }.$$
$$ [H_1,T^4_{\ 3}]=[H_1,T^3_{\ 4}]=[H_3,T^1_{\ 2}]=[H_3,T^2_{\ 1}]=0.$$
Using all the above algebra, we can find the other commutation relations of
the generators, which are shown in Appendix.
The above approach could be extended to obtain deformed $sl(n)$ ($n>4$)
algebra on the n-dim quantum space.

\vspace{0.8 cm}
\leftline{\large \bf 3. Conjugate spinor and deformed $su(4)$ algebra}
\vspace{0.8 cm}

Here we introduce another spinor $\overline x_i$ conjugate to the quantum
4-spinor $x^i$, in order to obtain deformed $su(m,n)$ ($m+n=4$) algebra.
The conjugate spinor is required to satisfy a commutation relation as
$$ \overline x_i \overline x_j=q^{-1} \overline x_j \overline x_i,
\quad (i<j). \eqno (3.1)$$
Further we define a commutation relation between $\overline x_i$ and $x^j$ as
follows,
$$ \overline x_i x^j = \hat R^{jk}_{\ \ i \ell}x^\ell \overline x_k.
\eqno (3.2)$$
This commutation relation is explicitly written as
$$ \overline x_i x^j = q^{-1} x^j \overline x_i, \quad (i \neq j),
\eqno (3.3)$$
$$ \overline x_i x^i = x^i \overline x_i + (1-q^{-2})\sum_{i<j} x^j
\overline x_j.$$
The above commutation relation has a center $c=x^i \overline x_i$.

Furthermore, we assume that the genarators act on $\overline x_i$ in the same
way as on $\partial_i$, (2.11),(2.15) and (A.1).
For example the generator $T^{\ell +1}_{\ \ell}$ acts on
$\overline x_{\ell +1}$ as
$$ T^{\ell +1}_{\ \ell} \overline x_{\ell +1} = q \overline x_{\ell +1}
T^{\ell +1}_{\ \ell} -q \overline x_{\ell}. \eqno (3.4)$$

We now consider conjugation between $x^i$ and $\overline x_i$.
The conjugation should be consistent with the commutation relations, (2.1),
(3.1) and (3.3).
We can find two types of consistent charge conjugations depending on values
of $q$.
If $q$ is real, the conjugation includes reversing the order, i.e.,
$\overline {ab}=\overline b \overline a$.
On the other hand, the conjugation for $|q|=1$ is simply written as
$\overline {ab}=\overline a \overline b$.
Under both conjugations, the spinor and conjugate spinor are related as
$$ \overline x_i = \eta^i (x^i)^*, \eqno (3.5)$$
where $*$ implies the complex conjugate and $\eta^i$ is a metric of the
$su(m,n)$ algebra.
The center $c$ is written in terms of $x^i$ and their complex conjugate as
$\eta^i x^i(x^i)^*$.
Thus, deformed $su(4)$ and $su(2,2)$ algebras have $\eta^i=(1,1,1,1)$ and
$(1,1,-1,-1,)$, respectively.

Now we consider charge conjugation of the genarators under the above
conjugations.
At first, we study the case with real $q$, where we also have to reverse the
order of the genarators and the spinors when taking the conjugation, i.e.,
$\overline {Tx}=\overline x \overline T$.
For example, we take the conjugation of the commutation relation between
$T^\ell_{\ \ell+1}$ and $x^{\ell +1}$ (2.6), so that we obtain
$$ \overline {T^\ell _{\ \ell+1}} \overline x_{\ell +1}=q
\overline x_{\ell +1}\overline {T^\ell_{\ \ell +1}}-q \eta^\ell
\eta^{\ell +1}\overline x_\ell. \eqno (3.6)$$
Comparing (3.6) with (3.4), we find charge conjugation of the generator
$T^\ell_{\ \ell +1}$ as follows,
$$ \overline {T^\ell_{\ \ell +1}} =\eta^\ell \eta^{\ell +1}
T^{\ell +1}_{\ \ell}. \eqno(3.7)$$
Similarly we find charge conjugation of the other generators as
$$ \overline {T^\ell _ {\ k}}= \eta^\ell \eta^k T^k_{\ \ell},\qquad
\overline {H_\ell}=H_\ell. \eqno (3.8)$$

On the other hand, in the case with $|q|=1$, we need not reverse the order,
i.e., $\overline {Tx}=\overline T \overline x$.
We take the conjugation of the commutaion realtion between
$T^\ell_{\ \ell +1}$ and $x$, so as to obtain
$$\overline{T^\ell _{\ \ell +1}}\overline x_{\ell +1}=q \overline x_{\ell +1}
\overline {T^\ell_{\ \ell +1}} + \eta^\ell \eta^{\ell +1} \overline x_\ell.
\eqno (3.9)$$
Thus we find the charge conjugation of ${T^\ell_{\ \ell+1}}$ as follows,
$$ \overline {T^\ell_{\ \ell+1}} = -q^{-1} \eta^\ell \eta^{\ell +1}
T^{\ell +1}_{\ \ell}.\eqno(3.10)$$
Similarly we obtain the charge conjugation of the other generators as,
$$ \overline {T^\ell_{\ k}}=-q^{2(\ell-k)+1} \eta^\ell \eta^k T^k_{\ \ell},
\qquad \overline H_\ell =- H_\ell. \eqno(3.11)$$

At last we obatain two types of deformed $su(m,n)$ ($m+n=4$) algebra as the
deformed $sl(4)$ algebra with two types of the conjugations, (3.8) and (3.11).
Especially, we obtain deformed conformal algebra represented on the quantum
4-spinor and the conjugate spinor.
Here, Lorentz generators are assigned to $T^1_{\ 2}$,$T^2_{\ 1}$,$H_1$,
$T^3_{\ 4}$,$T^4_{\ 3}$,$H_3$ and the other Cartan element $H_2$ corresponds
to degree of freedom of dilatation.
Further, the other generators of the $su(2,2)$ algebra correspond to linear
combinations of translation and conformal boost.

\vspace{0.8 cm}
\leftline{\large \bf 4. Quantum 6-vector}
\vspace{0.8 cm}

Classical algebras $su(4)$ and $su(2,2)$ are isomorphic to $so(6)$ and
$so(4,2)$, respectively.
In Ref. \cite{JO}, quantum generalization of the isomorphism between $sl(4)$
and $so(6)$ has been discussed.
Here we construct \lq \lq quantum 6-vector" from two copies of quantum
4-spinors in a similar way to Ref.\cite{SS,CSS-Watamura},
to represent the deformed conformal algebra on the quantum 6-vector.
In addition to $x^i$, the other quantum 4-spinor is denoted by $y^i$, which
satisfy the same commutation relation as one of $x^i$.
Further, the generators have the same actions on $y^i$ as those on $x^i$.
We assume that commutation relations between $x^i$ and $y^i$ are obtained as
$$ x^iy^j=\hat R^{ij}_{\ \ k\ell}y^kx^\ell.\eqno (4.1)$$
Eq. (4.1) is explicitly written as
$$ x^j y^i = q^{-1}y^i x^j, \qquad x^i y^i=y^ix^i,$$
$$ x^i y^j=q^{-1} y^j x^i +(1-q^{-2})y^ix^j, \eqno (4.2)$$
where $i<j$.

In the classical $su(4)$ algebra, a product of two fundamental quartet
representations, which are spinor representations of $so(6)$, is decomposed
into {\bf 10} and {\bf 6} representations.
The latter is a vector representation of $so(6)$ and antisymmetric for two
quartets.
Thus we need quantum generalization of the decomposition of the
representations, in order to construct the quantum 6-vector from the two
copies of the quantum 4-spinors.
A condition of the decomposition is closure of commutation relation algebra.
Namely, from linear combinations of $x^iy^j$, we have to pick up six elements
whose commutaion relations close themselves.
For that purpose, it is adequate to use the antisymmetric projection operaotr
${\cal A}$.
Commutation relations of ${\cal A}^{ij}_{\ \ k \ell}x^k y^\ell$ are close
themselves.
Six independet elements of ${\cal A}^{ij}_{\ \ k \ell}x^k y^\ell$ are
proportional to $x^i y^j-qx^jy^i$ ($i<j$) and hereafter we use
$$ a^{ij} \equiv x^i y^j-qx^jy^i, \qquad (i<j), \eqno (4.3)$$
as the quantum 6-vector instead of ${\cal A}^{ij}_{\ \ k \ell}x^k y^\ell$.
Actually, they satisfy commutation relations as follows,
$$ a^{ij}a^{ik}=q a^{ik}a^{ij},$$
$$ a^{ij} a^{jk} =q a^{jk} a^{ij},$$
$$ a^{ik} a^{jk} = q a^{jk} a^{ik},$$
$$ [a^{14},a^{23}]=0, \eqno (4.4)$$
$$ [a^{13},a^{24}]=(q-q^{-1})a^{14} a^{13},$$
$$ [a^{12},a^{34}]=q(-q+q^{-1})a^{14}a^{23} + (q - q^{-1}) a^{13} a^{24},$$
wher $i<j<k$.
The commutation relations are nothing but those of the quantum space for
$SO_q(6)$.
Further, this algebra has a center given by
$$ L \equiv qa^{14}a^{23}-a^{13}a^{24}+{1 \over q}a^{12}a^{34}. \eqno (4.5)$$
Similarly in terms of two copies of the quantum conjugate 4-spinors, we can
construct another quantum 6-vector ${\tilde a}$, which has a similar
commutation relation to (4.4).

Actions of the generators on the 6-vector are obtained through (2.11),(2.15)
and (A.1).
For example the actions of the Cartan elements on the 6-vector are obtained as
$$ [H_\ell, a^{\ell \ \ell +1}]=[H_\ell, a^{ij}]=0,$$
$$ H_\ell a^{\ell \ j}= q^2 a^{\ell \ i}H_\ell -q a^{\ell \  j},$$
$$ H_\ell a^{i \ \ell +1}=q^{-2}a^{i \  \ell +1}H_\ell +q^{-1} a^{i \ \ell +1}
, \eqno (4.6)$$
$$ H_\ell a^{i \ \ell}= q^2 a^{i \ \ell }H_\ell-qa^{i\ \ell},$$
$$ H_\ell a^{\ell+1 \ j}=q^{-2}a^{\ell+1 \ j} H_\ell + q^{-1}a^{\ell+1 \ j},$$
where $i<\ell$ and $\ell+1<j$, and $T^\ell_{\ k}$ ($\ell < k$) acts on the
6-vector as
$$ T^\ell_{\ k} a^{\ell k}=a^{\ell k} T^\ell_{\ k}+(1-q^2)
\sum^{k-1}_{m=\ell+1}a^{\ell m} T^\ell_{\ m},$$
$$ [T^\ell_{\ k}, a^{\ell j}]_q= [ T^\ell_{\ k}, a^{i \ell}]_q=0, \qquad
(k\neq j),$$
$$ T^\ell_{\ k}a^{ik}=q^{-1}a^{ik} T^\ell_{\ k}+a^{i \ell}+(q^{-1}-q)
\sum^{k-1}_{m=\ell+1}a^{i \ell} T^\ell_{\ m}, \qquad (i<\ell),$$
$$ T^\ell_{\ k} a^{kj}=q^{-1} a^{kj}  T^\ell_{\ k}+a^{\ell j}+(q^{-1}-q)
\sum^{k-1}_{m=\ell+1}a^{mj} T^\ell_{\ m},$$
$$  T^\ell_{\ \ell+2}a^{\ell+1 \ \ell+2}=q^{-1}a^{\ell+1 \ \ell +2}
T^\ell_{\ \ell+2}-qa^{\ell \ \ell+1}, \eqno (4.7)$$
$$ T^1_{\ 4} a^{24}=q^{-1}a^{24}T^1_{\ 4}+(q^{-1}-q)a^{23}T^1_{\ 3}-qa^{12},$$
$$ T^1_{\ 4} a^{34}=q^{-1}a^{34}T^1_{\ 4}+(q^2-1)a^{23}T^1_{\ 2}-qa^{13},$$
$$ [T^\ell_{\ k},a^{ij}]=0,$$
where in the last equation each index differs from the others.
Actions of  $T^k_{\ \ell}$ ($\ell<k$) on $a^{ij}$ are shown in Appendix.

In order to obtain four-dimensional space from six-dimensional one, we have
to compactify two coordinates.
In the classical coordinates, we can choose arbitrarily two compactified
coordinates, but in the quantum coordinates all compactification are not
consistent with commutation relations (4.4).
Note that commutation relations of $a^{13}$, $a^{14}$, $a^{23}$ and $a^{24}$
close themselves.
Thus, we should choose $a^{12}$ and $a^{34}$ as the compactified coordinates.
That is the remarkable fact in \lq \lq quantum compactification".

\vspace{1.8 cm}
\leftline{\large \bf 5. Conjugation and q-deformed conformal algebras}
\vspace{0.8 cm}

In this section we first consider another type of conjugation which
acts on 6-vectors as well as on the generators. Through this
conjugation procedure, the two sets of quantum 4-spinors are related
with each other.

As discussed in the previous section, the 6-vectors are represented as
tensor products of two quantum 4-spinors, $x$ and $y$ as
$$ a^{ij} \equiv x^i \otimes y^j-qx^j \otimes y^i, \qquad (i<j),
\eqno(5.1)$$
where we put $\otimes$ between $x$ and $y$ to show the tensor product
more explicitly.
We now introduce a new type of conjugation denoted by
${\bigcirc \hskip -.32cm * \hskip .17cm}$  in which we exchange the 4-spinors
$x$ and $y$
and subsequently take complex conjugates as follows:
$$(a^{ij})^{\conj} \equiv (y^j)^{*} \otimes (x^i)^{*}
-q^{*}(y^i)^{*} \otimes (x^j)^{*}, \qquad (i<j).
\eqno(5.2)$$
As we have mentioned in the previous section, a bi-spinor of two conjugate
4-spinors leads to another quantum 6-vector ${\tilde a}$ which can now be
identified with $a^{\conj}$ defined in the above equation.

Now it turns out that there exist two possibilities for this conjugation
in order for the six-vector to transform properly under the conjugation.
\vskip 1cm
\underline{i) Conjugation 1}

For real $q$, we take
$$
(x^i)^{*}=\eta^iy^{5-i} \qquad (i=1, \cdots, 4),
\eqno(5.3)$$
where $\eta^i=\pm 1$ depending upon metric for $su(m,n)$.  For $su(2,2)$ we
take $\eta^1=\eta^2=+1, \eta^3=\eta^4=-1$. This conjugation leads to the
reality condition:
$$
(a^{ij})^{\conj}=\eta^i\eta^{5-j}a^{5-j,5-i} \qquad (i<j).
\eqno(5.4)$$
Explicitly we have $(a^{14})^{\conj}=a^{14}$, $(a^{23})^{\conj}=a^{23}$,
$(a^{12})^{\conj}=\eta^1\eta^3a^{34}=-a^{34}$, $(a^{13})^{\conj}=\eta^1\eta^2
a^{24}=a^{24}$.

\underline{ii) Conjugation 2}

For $|q|=1$ or $q^*=1/q$, we set the following relation
$$
(x^i)^{*}=\eta^i\sqrt{q}y^i \ , \ (y^i)^{*}=\eta^i\sqrt{q}x^i
\qquad (i=1, \cdots, 4).
\eqno(5.5)$$
The conjugation of the 6-vector reads
$$
(a^{ij})^{\conj}=-\eta^i\eta^j a^{ij} \qquad (i<j),
\eqno(5.6)$$
namely we obtain the following reality condition:
$$\eqalign{
&(a^{12})^{\conj}=-a^{12} \quad , \quad
(a^{13})^{\conj}=a^{13} \quad , \quad
(a^{14})^{\conj}=a^{14} \cr
&(a^{23})^{\conj}=a^{23} \quad , \quad
(a^{24})^{\conj}=a^{24} \quad , \quad
(a^{34})^{\conj}=-a^{34}. \cr
}\eqno(5.7)$$

We now turn to the transformation of the generators under the conjugation
${\bigcirc \hskip -.32cm * \hskip .17cm}$.
The transformations are obtained in a procedure similar to the approach
discussed in section 3, (3.6) and (3.7).

For the conjugation 1, we find
$$\eqalign{
&(T^\ell_{\ \ell+1})^{\conj}=-q\eta^{\ell}\eta^{\ell+1}T^{\ell'}_{(\ell+1)'}
\qquad (\ell' \equiv 5-\ell) \cr
&(T^\ell_{\ \ell+2})^{\conj}=-q^3\eta^{\ell}\eta^{\ell+2}T^{\ell'}_{(\ell+2)'}
\quad , \quad (T^1_4)^{\conj}=-q^5\eta^1\eta^4T^4_1 \cr
&(H_1)^{\conj}=H_3 \quad , \quad (H_2)^{\conj}=H_2. \cr
}\eqno(5.8)$$
Or generally we have
$$
({T^i}_k)^{\conj}=-q^{2(k-i)-1}\eta^i\eta^k{T^{i'}}_{k'}.
\eqno(5.9)$$

For conjugation 2 we have
$$\eqalign{
\hskip 1.5 cm &(T^\ell_{\ \ell+1})^{\conj}=-{1\over q}\eta^{\ell}\eta^{\ell+1}
T^{\ell}_{(\ell+1)} \quad , \quad
({T^{\ \ell+1}}_\ell)^{\conj}=-q\eta^{\ell}\eta^{\ell+1}
{T^{(\ell+1)}}_{\ell} \cr
&(T^\ell_{\ \ell+2})^{\conj}=-q^{-3}\eta^{\ell}\eta^{\ell+2}
T^{\ell}_{(\ell+2)} \quad , \quad
({T^{\ \ell+2}}_\ell)^{\conj}=-q^3\eta^{\ell}\eta^{\ell+2}
{T^{(\ell+2)}}_{\ell} \cr
&({T^1}_4)^{\conj}=-q^{-5}\eta^1\eta^4{T^1}_4 \quad , \quad
({T^4}_1)^{\conj} = -q^5\eta^1\eta^4{T^4}_1 \cr
&(H_{\ell})^{\conj}=-H_{\ell}. \cr
}\eqno(5.10)$$
Generally we have
$$({T^i}_k)^{\conj}=-q^{2(i-k)+1}\eta^i\eta^k{T^i}_k.
\eqno(5.11)$$
The above conjugations are important in the case when we consider a
subalgebra of $su(2,2)$.
Namely, the ${\bigcirc \hskip -.32cm * \hskip .17cm}$-conjugations should
close in the subalgebra.

Let us now assign the $su_q(2,2)$ generators to the q-deformed conformal
algebra.
We first consider the case of conjugation 2.
Before going to the q-deformed case, we recapitulate the representation
for the the generators of classical $4D$ conformal group, namely translation
operators $P_{\mu}$, conformal boost operators $K_{\mu}$, Lorentz rotation
operators $M_{\mu\nu}$ and dilatation operator $D$ as $su(2,2)$ generators
(See e.g. ref.\cite{PVN}):

$$\eqalign{
&P_{\mu}=
\left(
\begin{array}{c|c}
0\  & \sigma_{\mu}\\
\hline
0\  & 0\  \\
\end{array}
\right)
\quad , \quad
K_{\mu}=
\left(
\begin{array}{c|c}
0\  & 0\  \\
\hline
{\bar \sigma}_{\mu} & 0\  \\
\end{array}
\right)
\cr
&M_{\mu\nu} =
\left(
\begin{array}{c|c}
\sigma_{\mu\nu}&0\ \\
\hline
0\ & {\bar \sigma}_{\mu\nu} \\
\end{array}
\right)
\quad , \quad
D =
\left(
\begin{array}{c|c}
-{1\over 2}{\bf 1} &0\ \\
\hline
0\ & {1\over 2}{\bf 1} \\
\end{array}
\right)
\cr
}\eqno(5.12)$$
where
$$\eqalign{
&\sigma_{\mu}=({\bf 1},\sigma^1,\sigma^2,\sigma^3)
\quad , \quad
{\bar \sigma}_{\mu}=({\bf 1},-\sigma^1,-\sigma^2,-\sigma^3) \cr
&\sigma_{\mu\nu}=\textstyle{{1\over 4}}(\sigma_{\mu}{\bar \sigma}_{\nu}-
\sigma_{\nu}{\bar \sigma}_{\mu}) \quad , \quad
{\bar \sigma}_{\mu\nu}=\textstyle{{1\over 4}}({\bar \sigma}_{\mu}\sigma_{\nu}-
{\bar \sigma}_{\nu}\sigma_{\mu}), \cr
}\eqno(5.13)$$
and $\sigma^i$ ($i=1,2,3$) are Pauli matrices.
For example $P_{\mu}$ and $K_{\mu}$ correspond to the generators
${T^i}_k$ as follows
$$\eqalign{
&P_0={T^1}_3 + {T^2}_4 \quad , \quad P_1={T^1}_4 + {T^2}_3 \cr
&P_2=-i{T^1}_4 + i{T^2}_3 \quad , \quad P_3={T^1}_3 - {T^2}_4 \cr
&K_0={T^3}_1 + {T^4}_2 \quad , \quad K_1=-{T^4}_1 - {T^3}_2 \cr
&K_2=-i{T^4}_1 + i{T^3}_2 \quad , \quad K_3=-{T^3}_1 + {T^4}_2. \cr
}\eqno(5.14)$$
For the Lorentz generators $M_{\mu\nu}$ and the dilatation operator
$D$ we take the following assignment:
$$\eqalign{
&M_{+}=M_{23}+iM_{31}=-i{T^1}_2-i{T^3}_4 \  , \
M_{-}=M_{23}-iM_{31}=-i{T^2}_1-i{T^4}_3 \cr
&M_3 = M_{21}=\textstyle{i\over 2}(H_1 + H_3) \cr
&L_{+}=M_{20}+iM_{01}=-i{T^1}_2+i{T^3}_4 \  , \
L_{-}=M_{20}-iM_{01}=i{T^2}_1-i{T^4}_3 \cr
&L_3 = M_{03} =\textstyle{1\over 2}(H_1 - H_3) \cr
&D=\textstyle{1\over 2}(H_1 + 2H_2 + H_3). \cr
}\eqno(5.15)$$
Now we take the same assignment of the generators for the q-deformed
case.  The q-deformed conformal algebra can be read off from the commutation
relations among the $sl_q(4,C)$ generators given in section 2 together with
Appendix.
What is remarkable here is that the Poincar{\'e} generators assigned above
form a closed subalgebra of the q-deformed conformal algebra in the case of
conjugation 2.

Now in the case of conjugation 1, we make the following assignment for
the generators.
For translation and conformal boost we take
$$\eqalign{
&{P^1}_3 \equiv {T^1}_3 + q^3{T^4}_2 \ , \
 {P^1}_4 \equiv {T^1}_4 + q^5{T^4}_1 \  , \
 {P^2}_3 \equiv {T^2}_3 + q{T^3}_2 \  , \
 {P^2}_4 \equiv {T^2}_4 + q^3{T^3}_1 \cr
&{K^1}_3 \equiv {T^1}_3 - q^3{T^4}_2 \  , \
 {K^1}_4 \equiv {T^1}_4 - q^5{T^4}_1 \  , \
 {K^2}_3 \equiv {T^2}_3 - q{T^3}_2 \  , \
 {K^2}_4 \equiv {T^2}_4 - q^3{T^3}_1 \cr
}\eqno(5.16)$$
and the same assignment for the Lorentz and the dilatational generators
as given in (5.15).  In this case, however, the q-deformed conformal algebra
does not include Poincar{\'e} algebra as a closed subalgebra. It turns out
that there is no possible contraction procedure which leads to the closed
Poincar{\'e} algebra.

\vspace{0.8 cm}
\leftline{\large \bf 6. q-deformed Poincar{\'e} algebra and Casimir invariant}
\vspace{0.8 cm}

Poincar{\'e} algebra, which is not a simple Lie algebra, cannot be deformed
in the standard prescription.  The conceivable method for obtaining
Poincar{\'e} algebra is through a contraction of anti-de Sitter algebra
$so(3,2)$ or conformal algebra $so(4,2)$ \break
\cite{Lukierski-Ruegg-Nowicki-Tolstoy,LN,LNR,CDPR}.
In this paper we extract it from the conformal algebra as a closed subalgebra.
Here we shall restrict ourselves to the case of conjugation 2.

For this choice we have q-deformed Lorentz algebra generated by
${T^1}_2$, ${T^2}_1$ , $H_1$ together with ${T^3}_4$, ${T^4}_3$ and $H_3$,
which are related to the conventional Lorentz generators as given in (5.15).
They satisfy the following commutation relation:
$$\eqalign{
&q^{-1}{T^{\ell+1}}_{\ell}{T^{\ell}}_{\ell+1}-q{T^{\ell}}_{\ell+1}
{T^{\ell+1}}_{\ell}= H_{\ell} \cr
&q^{-2}H_{\ell}{T^{\ell}}_{\ell+1} - q^2{T^{\ell}}_{\ell+1}H_{\ell}
= -(q+q^{-1}){T^{\ell}}_{\ell+1} \cr
&q^2H_{\ell}{T^{\ell+1}}_{\ell} - q^{-2}{T^{\ell+1}}_{\ell}H_{\ell}
= (q+q^{-1}){T^{\ell+1}}_{\ell} \quad (\ell=1,3). \cr
}\eqno(6.1)$$

{}From (5.14) we can form light-cone combinations which are associated with
single generators as
$$\eqalign{
&P_{\pm}=\textstyle{1\over 2}(P_0 \pm P_3) \quad , \quad
{\tilde P}_{\pm}=\textstyle{1\over 2}(P_1 \pm iP_2) \cr
& P_{+} = {T^1}_3 \quad , \quad P_{-} = {T^2}_4 \quad , \quad
{\tilde P}_{+} = {T^1}_4 \quad , \quad {\tilde P}_{-} = {T^2}_3 \cr
&K_{\pm}=\textstyle{1\over 2}(K_0 \pm K_3) \quad , \quad
{\tilde K}_{\pm}=\textstyle{1\over 2}(K_1 \pm iK_2) \cr
& K_{+} = {T^4}_2 \quad , \quad K_{-} = {T^3}_1 \quad , \quad
{\tilde K}_{+} = -{T^3}_2 \quad , \quad {\tilde K}_{-} = -{T^4}_1. \cr
}\eqno(6.2)$$

In this base the q-deformed commutation relations read
$$\eqalign{
&[{\tilde P}_{+},{\tilde P}_{-}] = 0 \quad , \quad
[P_{+},P_{-}] = -(q-q^{-1}){\tilde P}_{+}{\tilde P}_{-} \cr
&[{\tilde P}_{-}, P_{+}]_q = 0 \quad , \quad
[{\tilde P}_{+},P_{+}]_q = 0 \cr
&[P_{-}, {\tilde P}_{-}]_q = 0 \quad , \quad
[P_{-},{\tilde P}_{+}]_q = 0. \cr
}\eqno(6.3)$$

The commutation relations between the translational operators $P_{\mu}$ and
the
Lorentz generators $M_{\mu\nu}$ can be read off from the commutation
relations
of ${T^i}_k$ given in section 2 and Appendix.

Here we find the quadratic Casimir invariant for q-deformed Poincar{\'e}
group is given by
$$
C = q{T^1}_3{T^2}_4 - {T^1}_4{T^2}_3.
\eqno(6.4)$$
In terms of translational operators (5.15), the above Casimir can be written
as
$$
C=qP_{+}P_{-} - {\tilde P}_{+}{\tilde P}_{-},
\eqno(6.5)$$
which can be seen to commute with all the generators of the q-deformed
Poincar{\'e} algebra, and
$$4C= (P_0^2 -P_3^2-P_1^2-P_2^2)-{q\over 2}(q-q^{-1})(P_1^2+P_2^2)
+(q-1)(P_0^2-P_3^2).
\eqno(6.6)$$
corresponds to the q-deformed version of the relativistic mass squared
operator.
It would be interesting to investigate the q-deformed Klein-Gordon equation
for the q-deformed d'Alembertian corresponding to (6.6).

\vspace{2.8 cm}
\leftline{\large \bf 7. Concluding remarks}
\vspace{0.8 cm}

In this paper we investigated the q-deformed conformal and Poincar{\'e}
algebras in the framework of the quantum space for $sl_q(4,C)$.  We set up
the non-commutative relation for the quantum 4-spinors and analyzed the
differential calculus as well as the action of the generators on this quantum
space. Through the charge conjugations, we obtain two types of the deformed
$su(2,2)$ algebras and assign their generators to the elements of the
deformed conformal algebra.
The 6-vectors were constructed out of two sets of 4-spinors as tensor
products and their commutation relations were obtained.
We derived the q-deformed conformal algebra by setting up proper conjugations
of 4-spinors and 6-vectors.
The q-deformed Poincar{\'e} algebra was extracted as the closed subalgebra for
the suitable choice of the conjugation.

Now some remarks on the possible extentions are in order.
The present analysis of the one-parameter deformation could be extended to
multi-parameter deformations. In the present paper we studied the
deformation of the $4D$ conformal algebra corresponding to the deformed
$sl(4,C)$, which could be extended to higher dimensional algebras based on
the
present method.
It would be  extremely interesting to extend this formalism to $4D$
superconformal algebra on the quantum superspace\cite{KU}
which might shed some light on the quantum deformation of the
super-Poincar{\'e} algebra, which is now under investigation.

\vspace{0.8 cm}
\leftline{\large \bf Acknowledgement}
\vspace{0.8 cm}

The authors would like to thank P.~P.~Kulish, R.~Sasaki and C.~Schwiebert
for valuable discussions.

\newpage

\leftline{\large \bf Appendix}

In a similar way to (2.11), we find actions of the generators $T^k_{\ \ell}$
($\ell < k$) as follows,
$$T^k_{\ \ell } x^i=x^i T^k_{\ \ell } , \quad (i< \ell {\rm \ or \ } k<i),$$
$$T^k_{\ \ell } x^\ell = q x^\ell T^k_{\ \ell } + x^k,$$
$$T^k_{\ \ell } x^j = x^j T^k_{\ \ell } +(q^{-1}-q)x^k T^j_{\ \ell }, \quad
(\ell < j < k), $$
$$T^k_{\ \ell } x^k = q^{-1} x^k T^k_{\ \ell } , \eqno (A.1)$$
$$T^k_{\ \ell } \partial_\ell = q^{-1} \partial_\ell T^k_{\ \ell } ,$$
$$T^k_{\ \ell } \partial_k = q \partial_k T^k_{\ \ell } -q^{2(k-\ell)-1}
\partial_\ell + (q-q^{-1})\sum ^{k-1}_{j=\ell+1} q^{2(k-j)} \partial_j
T^j_{\ \ell } ,$$
$$T^k_{\ \ell }  \partial_i =\partial_i T^k_{\ \ell }, \quad (i\neq \ell,k).$$
Further, they satisfy commutation relations as
$$[T^{\ell}_{\ i},T^\ell_{\ j}]_q=[T^i_{\ \ell},T^j_{\ \ell}]_q=0,\qquad (i<j)
,$$
$$[T^4_{\ 3},T^2_{\ 1}]=[T^4_{\ 1},T^3_{\ 2}]=0,$$
$$[T^k_{\ j},T^j_{\ \ell}]_q=T^k_{\ \ell},\eqno (A.2)$$
$$[T^3_{\ 1},T^4_{\ 2}]=(q-q^{-1})T^4_{\ 1}T^3_{\ 2}.$$

In section 2, we have shown the basic commutation relations from which the
other commutaition relatins can be derived.
The remaining commutation relations are obtained as
$$ [H_\ell,T^\ell_{\ k}]_{q^2}=-qT^\ell_{\ k}, \qquad [H_{k-1},
T^\ell_{\ k}]_{q^2}=-qT^\ell_{\ k},\quad (k>\ell+1 {\rm \ or \ } k<\ell-1),$$
$$ [H_k,T^\ell_{\ k}]_{q^{-2}}=q^{-1}T^\ell_{\ k}, \qquad [H_{\ell-1},
T^\ell_{\ k}]_{q^{-2}}=q^{-1}T^\ell_{\ k},\quad (k>\ell+1 {\rm \ or \ }
k<\ell-1),$$
$$ [H_2,T^1_{\ 4}]=[H_2,T^4_{\ 1}]=0,$$
$$[T^\ell_{\ j},T^{\ell +1}_{\ \ell}]_{q^{-1}}=(1-q^{-2})T^{\ell+1}_{\ j}
H_\ell-q^{-1}T^{\ell+1}_{\ j}, \quad (\ell+1<j)$$
$$[T^\ell_{\ \ell+1},T^j_{\ \ell}]_{q^{-1}}=(q^2-1)T^j_{\ \ell+1}H_\ell
-qT^j_{\ \ell+1}, \quad (\ell+1<j)$$
$$[T^i_{\ k},T^k_{\ j}]_{q^{-1}}=T^i_{\ j}, \quad (i<j,j=k-1) {\rm \ or \ }
(j<i,i=k-1),$$
$$[T^\ell_{\ 4},T^\ell_{\ 1}]_q=[T^1_{\ \ell},T^4_{\ \ell}]_q=0, \quad
(\ell=2,3),$$
$$[T^1_{\ 4},T^3_{\ 2}]=[T^1_{\ 3},T^4_{\ 2}]=0,$$
$$[T^2_{\ 3},T^4_{\ 1}]=[T^2_{\ 4},T^3_{\ 1}]=0,\eqno (A.3)$$
$$[T^i_{\ 4},T^4_{\ j}]_{q^{-1}}=T^i_{\ j}-\lambda T^3_{\ j}T^i_{\ 3},\quad
((i,j)=(1,2) {\rm \ or \ } (2,1)),$$
$$\eqalign{[T^1_{\ i},T^j_{\ 1}]_{q^{-1}}= & -q^{j-i+2}T^j_{\ i}+\lambda
T^j_{\ 2}T^2_{\ i}+\lambda q^{j-i+2}T^j_{\ i}(H_1+H_2) \cr
& -\lambda^2T^j_{\ 2}T^2_{\ i}H_1 -\lambda^2q^{j-i+2}T^j_{\ i}H_1H_2,\cr
& \hskip 5cm ((i,j)=(3,4) {\rm \ or \ } (4,3)),}$$
$$\eqalign{[T^\ell_{\ \ell+2},T^{\ell+2}_{\ \ell}]_{q^{-2}}= & -q^{-1}H_\ell-
qH_{\ell+1}-q^{-1}\lambda T^{\ell+1}_{\ \ell}T^\ell_{\ \ell +1}+q^{-1}\lambda
T^{\ell+2}_{\ \ell+1}T^{\ell+1}_{\ \ell+2} \cr
& +q^{-1}\lambda H_\ell H_{\ell+1}-q\lambda^2T^{\ell+1}_{\  \ell+2}
T^{\ell+2}_{\ \ell+1}H_{\ell},}$$
$$\eqalign{[T^1_{\ 4},T^4_{\ 1}]_{q^{-2}}=-qH_1-q^{-1}(2-q^2)H_2-q(2-q^2)H_3
-q^{-1}\lambda T^4_{\ 2}T^2_{\ 4}+q\lambda T^1_{\ 2}T^2_{\ 1} \cr
-q^{-1}\lambda T^3_{\ 1}T^1_{\ 3}+q^{-1}(2-q^2)\lambda T^4_{\ 3}T^3_{\ 4}+
(1+q^{-2})\lambda T^4_{\ 2}T^2_{\ 4}H_1 -\lambda qH_1H_2 \cr
-\lambda q^3 H_1 H_3+\lambda q^{-1} (2-q^{-2}) H_2 H_3-(q+q^{-1})
\lambda^2T^3_{\ 2}T^2_{\ 3}H_1+\lambda^2 T^3_{\ 2} T^2_{\ 3} \cr
-q^{-1}(2-q^2)\lambda^2T^3_{\ 4}T^4_{\ 3}H_2+q\lambda^2 T^4_{\ 3}T^3_{\ 4}H_1
-q^3\lambda^3T^3_{\ 4} T^4_{\ 3}H_1H_2,}$$
where $\lambda=q-1/q$.
At last, actions of $T^k_{\ \ell}$ ($\ell<k$) on the 6-vector are obtained as
follows
$$ T^k_{\ \ell} a^{\ell k}=a^{\ell k} T^k_{\ \ell},$$
$$ T^k_{\ \ell}a^{\ell j}=qa^{\ell j} T^k_{\ \ell}+a^{k j},\quad (k<j),$$
$$ T^k_{\ \ell} a^{ik}=q^{-1}a^{ik}T^{k}_{\ \ell}, \quad (i \neq \ell), $$
$$ T^k_{\ \ell} a^{i \ell}=q a^{i \ell}T^k_{\ \ell} +a^{ik},$$
$$ T^k_{\ \ell} a^{kj}=q^{-1}a^{kj}T^k_{\ \ell} ,\eqno (A.4)$$
$$ T^k_{\ \ell} a^{\ell j}=q a^{\ell j} T^k_{\ \ell} +(1-q^2)a^{\ell k}
T^j_{\ \ell}-qa^{j k}, \quad (\ell< j <k),$$
$$T^4_{\ 2}a^{13}=a^{13}T^4_{\ 2}-\lambda a^{14}T^3_{\ 2},$$
$$T^3_{\ 1}a^{24}=a^{24}T^3_{\ 1}-\lambda a^{34}T^2_{\ 1},$$
$$T^4_{\ 1}a^{23}=a^{23}T^4_{\ 1}-\lambda a^{24}T^3_{\ 1}+q\lambda a^{34}
T^2_{\ 1},$$
$$T^k_{\ \ell}a^{ij}=a^{ij}T^k_{\ \ell}, \quad (k<i {\rm \ or \ } j<\ell).$$


\end{document}